\documentclass[3p,times,authoryear]{elsarticle}
\usepackage{ecrc}

\volume{00}

\firstpage{1}

\journalname{Computer Communications}

\runauth{}

\usepackage{amssymb}
\usepackage{algorithm}
\usepackage{amsthm}
\usepackage{algorithmicx}
\usepackage{algpseudocode}
\usepackage{multirow}
\usepackage{amsmath}
\usepackage[figuresright]{rotating}
\usepackage{graphicx}%插入图片
\usepackage{subfigure}
\usepackage{float}
\usepackage{epstopdf}
\usepackage{color}

\newcounter{RomanNumber}
\newcommand{\MyRoman}[1]{\setcounter{RomanNumber}{#1}\Roman{RomanNumber}}
\usepackage[colorlinks, citecolor=blue]{hyperref}
\begin{document}

\begin{frontmatter}
%%\hypersetup{CJKbookmarks=true}


\dochead{}

\title{Privacy-preserving Double Auction Mechanism Based on Homomorphic Encryption and Sorting Networks}

\author[Address2]{Yin Xu}
\ead{15215603225@163.com}

%%作者的\corref{lable}就是在脚注，用的是*，第n个有n*
%%\fnref{lable} 脚注，这个才是真的，但是没显示
%%\author[lindexiAddress]{lindexi\_gd \corref{cor1} \fnref{fn1} }
%%地址写在下面\address[lindexiAddress]{The lindexi's address }
%%地址是a,b,c  label不区分大小写

\author[Address2]{Zhili Chen \corref{cor1}}
\ead{zlchen3@ustc.edu.cn}

\author[Address2]{Hong Zhong}
\ead{zhongh@mail.ustc.edu.cn}

\cortext[cor1]{Corresponding author}
%\cortext[cor2]{Principal corresponding author}

%\fntext[fn1]{This is the specimen author footnote.}
%\fntext[fn2]{Another author footnote, but a little more longer.}
%\fntext[fn3]{Yet another author footnote. Indeed, you can have
%    any number of author footnotes.}

%\address[Address1]{School of Anhui University, Hefei, China}

\address[Address2]{School of Computer Science and Technology, Anhui University, Hefei, China}

\begin{abstract}
As an effective resource allocation approach, double auctions (DAs) have been extensively studied in electronic commerce. Most previous studies have focused on how to design strategy-proof DA mechanisms, while not much research effort has been done concerning privacy and security issues. However, security, especially privacy issues have become such a public concern that the European governments lay down the law to enforce the privacy guarantees recently. In this paper, to address the privacy issue in electronic auctions, we concentrate on how to design a privacy-preserving mechanism for double auctions by employing Goldwasser-Micali homomorphic encryption and sorting networks. We achieve provable privacy such that the auctions do not reveal any bid information except the auction results, resulting in a strict privacy guarantee. Moreover, to achieve practical system performance, we compare different sorting algorithms, and suggest using the faster ones. Experimental results show that different sorting algorithms may have great effect on the performance of our mechanism, and demonstrate the practicality of our protocol for real-world applications in electronic commerce.
\end{abstract}


\end{frontmatter}


\section{Introduction}
With the rapid development of electronic commerce over the Internet, a large number of auction services exist on the Internet during the past a few years. Double auctions are adopted by many of current electronic exchanges for matching buyers and sellers (\citealp{Shi2016An}). Other auctions are also widely employed as efficient mechanisms to allocate goods, resource, etc., to promote the ideality of competitive pricing and to maximize the commodity value. The rapidity and convenience of the Internet has accelerated online auctions to individuals, businesses and governments around the world, admittedly, and the range of items sold by auctions also has been greatly extended by e-commerce, where millions of transactions can be accomplished in the blink of an eye. For example, eBay-the most popular and well-known auction site on the Internet announced that its total commodity trading in 2017 was US\$88.4 billion, with net revenues of US\$9.567 billion, an increase of 7\% over the 2016 (\url{http://www.ebay.com}).

At present, there are four basic types of auctions: the English (also called the oral, open, or ascending-bid) auction, the Dutch (or descending-bid) auction, the sealed-bid auction, and the double auction (DA) (\citealp{Mcafee1992A}). In different application scenarios, each type of auction has its advantages and disadvantages. Numerous auction mechanisms have been proposed in order to satisfy various economic properties such as truthfulness, profit maximization, and social efficiency (cf. \citealp{Iosifidis2015A}; \citealp{Li2016A}; \citealp{Cai2014On} and references therein). Unfortunately, living in the era of information, it is extremely difficult to guarantee complete security and privacy. According to annual reports of the National White Collar Crime Center (NW3C), online auctions constitute the main access point for criminals to acquire identity information and strip consumers of money or merchandise (NW3C 2014). Once a large amount of bid information is leaked, it will bring irreparable losses to the buyers or the sellers. For instance, in 2014 the famous online auction giant eBay was hacked, and quantities of auction data were stolen, which brought a net loss of US\$18 million to the company in the first quarter of that year.

There are two major research directions in regard to auctions. One is the research into auction protocols. An auction protocol defines the valid behaviors of participants in order to ensure the efficiency and privacy. The other research direction is on bidding strategies in terms of cost saving, perceived bidder enjoyment and bidder satisfaction. Our paper lays more emphasis on the former.

\subsection{The problem of bid privacy disclosure}
It is well acknowledged that information privacy has become an issue of universal concern recently. Auctions are not immune to privacy disclosure and this commitment to a correct process can be greatly difficult to achieve. There are many reasons for the cause of this phenomenon, such as the auctioneer breaks the rules of the auction in favor of some bidders, or the collusion between the auctioneer and a part of bidders to share a large extra profit, or criminals steal bid information for money. More than ethically troubling, privacy disclosure is undesirable because it can bring about an efficiency loss, a money loss, unfair distribution and malignant competition. Privacy disclosure is a widespread, real-world and noteworthy problem, as illustrated by the following examples:
\begin{itemize}
\item{The Attorney General of New York, Eliot Spitzer, accused major insurance brokers in the U.S. These insurance brokers manipulated bids in order to assign contracts to bidders in exchange for bribes which were up to \$845m just in the year 2003 (\citealp{lengwiler2005bid}).}
\item{In May 2016, a Russian hacker stole 272.3 million e-mail messages, including 40 million Yahoo mails, 33 million Microsoft mails and 24 million Google mails. Later, this information flowed into the black market in Russia and sold for less than US\$1.}
\item{In 2017, Bithumb, South Korea's first virtual currency trading platform, was hacked and about 30,000 personal user data were stolen. After the information was leaked, the hacker used personal user data to steal the money in the account and telephoned the exchange staff to make phone fraud.}
\end{itemize}\par
\indent When bid is the only factor in determining the winner of an auction, then an increasing number of participants tend to adopt a privacy-preserving, sealed-bid auction to prevent bid privacy disclosure.

McAfee's double auction is most widely used because of its economic robustness such as truthfulness, individual rationality and ex-post budget balance (\citealp{Chen2013PS}). However, most studies on it have focused on how to design strategy-proof DA mechanisms (\citealp{Colini2016Approximately}; \citealp{Nassiri2015Bidding}; \citealp{Samimi2016A}; \citealp{Shi2016An}; \citealp{Ma2007An}), while little research effort has been done concerning privacy and security issues in the auctions, leading to bid privacy disclosure largely. This would in turn lead to a series of severe consequences.
\begin{itemize}
\item{For the sellers, it can easily adjust its bid according to the buyers' and other sellers' bids to obtain extra profit.}
\item{For the buyers, similarly, it can easily adjust its bid according to the sellers' and other buyers' bids to lower its cost.}
\item{For the auctioneer, since it can obtain bid information directly from both sellers and buyers, it may tamper with the auction result, and hence destroy the truthfulness of the whole auction.}
\end{itemize}\par

\subsection{Our solution}
\indent Recently, some mechanisms for privacy-preserving auctions have been proposed (\citealp{Huang2013SPRING}; \citealp{Nojoumian2014Efficient}). These works dealt with only single-sided auctions. For double auctions, there are multiple potential buyers and potential sellers simultaneously permitted to bid to exchange designated goods, and a more complex logic is needed to match between buying and selling requirements. Indeed, there have also been some double auction mechanisms with privacy protection, such as \citet{Li2015Secure} and \citet{Chen2015ITSEC}. However, they are mainly focused on the application setting of double spectrum auctions, and use quite different security techniques such as garbled circuits and secret sharing. In contrast, this paper will employ Goldwasser-Micali homomorphic encryption to provide an alternative privacy-preserving mechanism based on McAfee's double auction mechanism, and experimentally validate that different sorting algorithms may have great influence on the performance of the proposed mechanism.

\begin{figure}[!htb]
    \centering
	\includegraphics{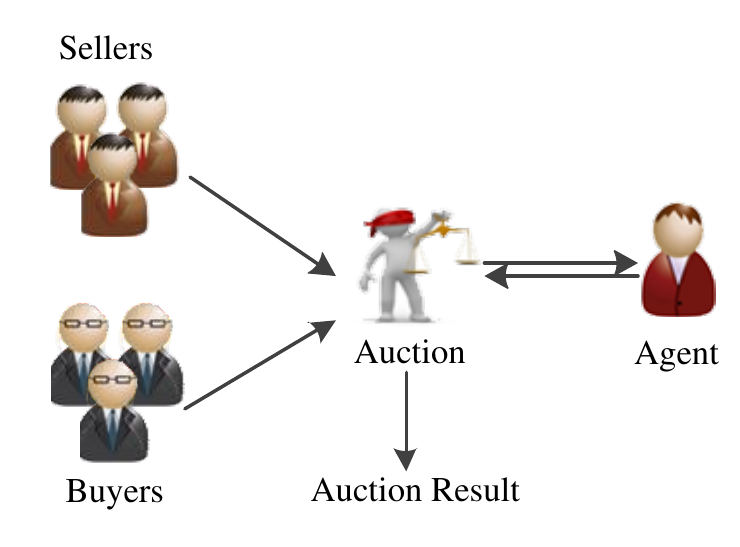}
	\caption{Privacy-preserving double auction architecture}
	\label{fig1}
\end{figure}

Architecturally, our auction framework is depicted in Fig. \ref{fig1}: each buyer (or seller) submits his bid to the auctioneer. But both buyers and sellers do not reveal their respective bid information during the auction. Instead, they submit encrypted bids to the auctioneer who determines the winners and assignments in company with the agent, which plays a role in administering the private key for the privacy-preserving auction. As long as the auctioneer and the agent do not collude to manipulate the auction, our mechanism can achieve provable privacy. Our contributions can be summarized as follows.
\begin{enumerate}[\indent (1)]
\item As far as we know, we are the first to design a privacy-preserving double auction with Goldwasser-Micali homomorphic encryption, and implement the mechanism with different sorting algorithms.
\item We theoretically prove the privacy of our double auction mechanism, with the privacy formulation from secure two party computations.
\item We experimentally evaluate and compare the performance of our mechanism using different sorting algorithms.
\end{enumerate}\par
\indent The remainder of this paper is organized as follows. Section 2 introduces some technical preliminaries. We present our mechanism in detail, and theoretically prove its privacy in Section 3. In section 4, we implement our mechanism and evaluate its running performance with different sorting algorithms. Section 5 reviews related work and discusses the difference about those studies briefly. Finally, we make a conclusion in Section 6.

\section{TECHNICAL PRELIMINARIES}
\indent In this section, we introduce some technical preliminaries for understanding our mechanism subsequently.
\subsection{McAfee's Double Auction}
\indent We assume that there are $M$ sellers and $N$ buyers in the auction, and all auction goods are homogeneous which means that each individual good is identical and that all goods have little or no differentiation in terms of features, benefits, or quality. Each seller $s_i$ bids $V_i^s$ to sell a good, and each buyer $b_i$ bids $V_i^b$ to buy a good. The auction proceeds as follows:
\begin{enumerate}[\indent (1)]
\item Bid sorting: Sort buyers' bids in non-increasing order and sellers' bids in non-descending order:
\[ V_1^b\geq V_2^b\geq \cdots \geq V_N^b
\]
\[ V_1^s\leq V_2^s\leq \cdots \leq V_M^s
\]
\item Winner determination: Find the index of the last profitable transaction $k$ shown in Equation (1) by means of comparing $V_i^b$ and $V_i^s$ with $i$ starting from 1. Then the first $(k-1)$ sellers and the first $(k-1)$ buyers are the auction winners.
\begin{equation}
k=argmax \{ V_i^s\leq V_i^b\}
\end{equation}
\item Pricing: Pay each winning seller equally by $V_k^s$, and charge each winning buyer equally by $V_k^b$.
\end{enumerate}\par

\subsection{Homomorphic Encryption}
\indent In our design of a privacy-preserving auction mechanism, a semantically secure cryptosystem is needed. Homomorphic encryption is a special type of encryption technique that allows computations to be done on encrypted data. In our paper, we choose the Goldwasser-Micali cryptosystem $(K,E,D)$ (\citealp{Blum1985An}), where $K$, $E$ and $D$ denote the key generation algorithm, encryption algorithm, and decryption algorithm, respectively. These algorithms of Goldwasser-Micali scheme definition can be described as follows.
\begin{itemize}
\item{\emph {Key generation}: the algorithm takes a security parameter as input and then generates two distinct large prime numbers $p$ and $q$, computes $n=pq$ and finds a non-residue $x$. The public key $k_p$ is $(x,n)$, and the private key $k_s$ is $(p,q)$.}
\item{\emph {Encryption algorithm}: a message $m\in \{0,1\}^l$ is encrypted as the ciphertext $c$ which makes use of the public key:}
\[
c=y^2x^m(mod\ n)
\]
Where $y$ is randomly chosen from $\mathbb {Z}^*_n$.
\item{\emph {Decryption algorithm}:	a ciphertext $c$ is decrypted as the message $m$ using the prime factorization $(p,q)$, where $m=0$ if $c$ is a quadratic residue, $m=1$ otherwise.}
\end{itemize}\par
\indent This cryptosystem possesses the XOR homomorphism, that is, for any $m,m'\in \{0,1\}$ the following equation holds.
\begin{equation}
D(E(m,k_p)\times E(m',k_p),k_s)=m\oplus m'
\end{equation}
\indent In particular, a ciphertext $c=E(m,k_p)$ can be randomized into other forms without changing its corresponding plaintext, as follows.
\begin{equation}
c'=E(m,k_p )\times E(0,k_p )
\end{equation}

\subsection{Cryptographical Privacy}
\indent In our auction context, the standard definition of cryptographical privacy against semi-honest adversaries in two-party computations can be described as follows (\citealp{Chen2016On}).
\newdefinition{rmk}{Definition}
\begin{rmk}[Privacy against semi-honest attackers]Let $f(x,y)$ be a functionality with two inputs $x$ and $y$, and two outputs $f^A (x,y)$ and $f^B (x,y)$. Suppose that protocol \MyRoman{2} computes functionality $f(x,y)$ between two parties Alice and Bob. Let $V_A^{\MyRoman{2}}(x,y)$ (resp. $V_B^{\MyRoman{2}}(x,y))$ represent Alice's (resp. Bob's) view during an execution of \MyRoman{2} on $(x,y)$. In other words, if $(x,r_A^{\MyRoman{2}})$ (resp. $(y,r_B^{\MyRoman{2}}))$ denotes Alice's (Bob's) input and randomness, then
\[
V_A^{\MyRoman{2}}(x,y)=(x,r_A^{\MyRoman{2}},m_1,m_2,\cdots,m_t)
\]
\[
V_B^{\MyRoman{2}}(x,y)=(y,r_B^{\MyRoman{2}},m_1,m_2,\cdots,m_t)
\]\par
Where $\{m_i\}$ denote the messages passed between the parties.
\\ \indent Let $O_A^{\MyRoman{2}}$ (resp. $O_B^{\MyRoman{2}}$) denote Alice's (Bob's) output after an execution of {\MyRoman{2}} on $(x,y)$, and $O^{\MyRoman{2}} (x,y)=(O_A^{\MyRoman{2}} (x,y),O_B^{\MyRoman{2}} (x,y))$. Then we say that protocol {\MyRoman{2}} preserves privacy (or is secure) against semi-honest adversaries if there exist probabilistic polynomial time (PPT) simulators $S_1$ and $S_2$ such that
\begin{equation}
\{(S_1 (x,f_A (x,y)),f(x,y))\}\equiv \{(V_A^{\MyRoman{2}} (x,y),O^{\MyRoman{2}} (x,y))\}
\end{equation}
\begin{equation}
\{(S_2 (y,f_B (x,y)),f(x,y))\}\equiv \{(V_B^{\MyRoman{2}} (x,y),O^{\MyRoman{2}} (x,y))\}
\end{equation}\par
Where $\equiv$ denotes computational indistinguishability.
\end{rmk}\par
\indent In the sense of cryptography, to protect both sellers' and buyers' bids (privacy), the secure protocol should reveal nothing about these bids except what is revealed from the auction outcome.

\section{OUR MECHANISM}
\indent In this section, we propose a privacy-preserving auction mechanism, whose auction framework is illustrated in Fig.~\ref{fig1}. As shown in the figure, the auctioneer receives the (encrypted) bids from all buyers and sellers and then performs the privacy-preserving auction, while the agent is in charge of assisting the auctioneer in the process of auction. Adopting the above framework, we design a privacy-preserving auction mechanism. Note that in the paper, we assume that the auctioneer and the agent do not collude with each other. This assumption is essentially necessary to achieve cryptographical privacy, since if the auctioneer and the agent collude with each other, namely, both computation parties are corrupted, then all bid information will be certainly leaked to the adversary.

\begin{rmk}[Encrypted Bit Vector]The Encrypted Bit Vector (EBV) representation of value $v$ is a vector $E^L (V)$ of ciphertexts like
\[
E^L (V)=(c_L,c_{L-1},\cdots ,c_1)=(E(\alpha _L ),E(\alpha_{L-1}),\cdots,E(\alpha_1))
\]\par
Where $E(.)$ is Goldwasser-Micali's encryption function, $L$ is the bit length, $(\alpha_L,\alpha_{L-1},…,\alpha_1 )$  denotes the binary representation of $V$, with $\alpha_L$ the most significant bit, and $\alpha_1$ the least significant bit.
\end{rmk}
\subsection{Privacy-preserving Auction Framework}
In order to guarantee that neither the auctioneer nor the agent can obtain anything about bid information except the final auction outcome during the auction, we design our privacy-preserving double auction mechanism using homomorphic encryption and sorting networks in Protocol \ref{Protocol1}. In our paper, we assume that there is an authenticated and secure communication channel between buyers (or sellers) and the auctioneer, and between the auctioneer and the agent. The design rationale can be described as follows.
\\ \indent \textbf{(1) Setup}
\\ \indent The agent generates its secret-public key pair of Goldwasser-Micali homomorphic encryption, and then publishes its public key to all the participants of the auction, including the auctioneer, the sellers and the buyers.
\\ \indent \textbf{(2) Bidding}
\\ \indent Each buyer or seller encrypts its bid-ID pair in the form of EBV using the public key of the agent, and submits its EBV pair to the auctioneer. Note that, in order to determine privately who the winners are, we need to encrypt the buyers' or sellers' IDs in the EBV form together with their bids. Otherwise, the bid ranking information of both sellers and buyers would simply leak in the next step.
\\ \indent \textbf{(3) Private sorting}
\\ \indent The auctioneer and the agent cooperate to privately sort the buyers' EBV bid-ID pairs in non-ascending order of bids, and the sellers' EBV bid-ID pairs in non-descending order of bids, making use of the XOR homomorphism of the Goldwasser-Micali cryptosystem. How to privately perform the sorting procedure is detailed in Section 3.4. In this process, the auctioneer knows nothing about the buyers' or sellers' bids; similarly, the agent only plays a role in providing some necessary calculation results to the auctioneer and also knows nothing about the bids. In the end, the auctioneer holds the private sorting for both the sellers' and the buyers' bid-ID pairs.
\\ \indent \textbf{(4) Private winner determination}
\\ \indent The main goal of this step is to obtain the encrypted auction outcome under the collaborative work of the auctioneer and the agent. A natural procedure for winner determination proceeds as follows. In order to determine the index of the last profitable transaction, the auctioneer compares the sorted seller's EBV bids and buyer's EBV bids one by one, which should start from the minimum EBV bid in the seller set and the maximum EBV bid in the buyer set. And then the encrypted comparison results are sent to the agent, who decrypts these encrypted data using the private key, and sends the index of the last profitable transaction $k$ back to the auctioneer. Getting $k$, the auctioneer can easily obtain the EBV auction outcome.
\\ \indent \textbf{(5) Outcome release}
\\ \indent A naive idea to release the auction outcome is as follows. The auctioneer sends the EBV IDs of the winners, namely the first $(k-1)$ sellers and the first $(k-1)$ buyers, and the EBV clearing prices $E^L(V_k^s)$ and $E^L(V_k^b)$ to the agent. The agent decrypts all these received EBV data, and sends back the plain final auction outcome including the IDs of winners and the clearing prices to the auctioneer.
\\ \indent The idea above seems to work well: the auctioneer and the agent cooperate to determine the winners and no exact bids are leaked to them. However, there is indeed some information about the bids leaking. It is noteworthy that each buyer's or seller's ID and bid are combined together as a whole, to put it another way, each ID is uniquely mapped to a bid. So, if the encrypted IDs of winners are decrypted just in the current orders of sellers and buyers, which are sorted in term of bids, the bid ranking information of both seller and buyer winners would be certainly leaked. The leaked information is obviously more than what we can infer from the auction outcome. Thus, in the sense of cryptography, the above procedure is not really secure.
\floatname{algorithm}{Protocol}
\begin{algorithm}[H]
  \caption{ Privacy-preserving Double Auction}
  \label{Protocol1}
  \begin{algorithmic}
    \Ensure   Double Auction results in the clear
    \State {\bfseries Step1: Setup}
    \State \indent The agent generates its secret-public key pair $(sk,pk)$ of homomorphic encryption, and then publishes its public key $pk$ to the auctioneer, the buyers and the sellers.
    \State {\bfseries Step2: Bidding}
    \State (1) Each seller sends its encrypted bid-ID pair $(E^L_{pk}(V_i^s),E^{L'}_{pk}(s_i))$ to the auctioneer.
    \State (2) Each buyer sends its encrypted bid-ID pair $(E^L_{pk}(V_i^b),E^{L'}_{pk}(b_i))$ to the auctioneer.
    \State {\bfseries Step3: Private sorting}
    \State \indent The auctioneer and the agent cooperate to privately sort sellers' encrypted bid-ID pairs in non-descending order, and buyers' encrypted bid-ID pairs in non-ascending order, both in term of bids:
    \State \centerline{\large{$(E^L_{pk}(V_1^s),E^{L'}_{pk}(s_1)),(E^L_{pk}(V_2^s),E^{L'}_{pk}(s_2)),\cdots, (E^L_{pk}(V_M^s),E^{L'}_{pk}(s_M))$}}
    \State \centerline{\large{s.t. $ E^L_{pk}(V_{1}^{s})\leq E^L_{pk}(V_{2}^{s})\leq \cdots\leq E^L_{pk}(V_{M}^{s})$}}
    \State \centerline{\large{$(E^L_{pk}(V_1^b),E^{L'}_{pk}(b_1)),(E^L_{pk}(V_2^b),E^{L'}_{pk}(b_2)),\cdots, (E^L_{pk}(V_N^b),E^{L'}_{pk}(b_N))$}}
    \State \centerline{\large{s.t. $ E^L_{pk}(V_{1}^{b})\geq E^L_{pk}(V_{2}^{b})\geq \cdots\geq E^L_{pk}(V_{N}^{b})$}}
    \State \indent Note that, for convenience, here we abuse the same indexes after sorting.
    \State {\bfseries Step4: Private winner determination}
    \State \indent The auctioneer and the agent cooperate to privately compare $E^L_{pk}(V_i^s)$ with $E^L_{pk}(V_i^b)$, until the index of the last profitable transaction $k$ is found.
    \State \centerline{\large{$ k=argmax \{ E^L_{pk}(V_k^s )\leq E^L_{pk}(V_k^b )\}$}}
    \State \indent The auctioneer gets the encrypted auction outcome including the encrypted IDs of the seller and buyer winners, and the encrypted clearing prices, as follows.
    \State \indent Encrypted winner IDs:
    \begin{equation}
    E^{L'}_{pk}(s_1),E^{L'}_{pk}(s_2),\cdots,E^{L'}_{pk}(s_{k-1})
    \end{equation}
    \begin{equation}
    E^{L'}_{pk}(b_1),E^{L'}_{pk}(b_2),\cdots,E^{L'}_{pk}(b_{k-1})
    \end{equation}
    \State \indent Encrypted clearing prices: $E^L_{pk}(V_k^s )$ and $E_{pk}^L(V_k^b )$.
    \State {\bfseries Step5: Outcome release}
    \State \indent The auctioneer randomly permutes the two arrays (6) and (7), respectively, and then sends the resulted arrays to the agent.
    \State \indent The agent decrypts the encrypted auction outcome:
    \State \centerline{\large{$D^L_{sk}(E^L_{pk}(V^s_k))=V^s_k,D^{L'}_{sk}(E^{L'}_{pk}(s_i))=s_i(i=1,2,...,{k-1})$}}
    \State \centerline{\large{$D^L_{sk}(E^L_{pk}(V^b_k))=V^b_k,D^{L'}_{sk}(E^{L'}_{pk}(b_i))=b_i(i=1,2,...,{k-1})$}}
    \State \indent The agent then sorts the two arrays of $s_i$'s and $b_i$'s in ascending order, and returns all decrypted auction outcome to the auctioneer.

    Finally, the auctioneer charges the expenses from each winning buyer and pays the earnings to each winning seller.
  \end{algorithmic}
\end{algorithm}

In order to make this natural procedure of winner determination secure, something has to be done before the decryption of the auction outcome, so that neither the auctioneer nor the agent can match the ID of any winner to its bid ranking position. Specifically, we let the auctioneer randomly permute the encrypted IDs of both seller and buyer winners before sending to the agent for decryption, and let the agent also randomly permute the decrypted IDs of both seller and buyer winners before returning back to the auctioneer (In protocol \ref{Protocol1}, instead of applying a random permutation, we sort IDs in ascending order, assuming that the bids are independent on IDs.). Finally, when the auction is over, the auctioneer announces final winner list and charges the expenses from each winning buyer, and pays the earnings to each winning seller.
\\ \indent The global algorithm is illustrated in Protocol \ref{Protocol1}.

\subsection{Private Gate Computation}
\indent In term of computation theory, any computable function can be represented by a Boolean circuit with only XOR and AND gates. Thus, to compute an auction privately, we need to compute both XOR and AND gates privately. With the XOR homomorphism of the Goldwasser-Micali cryptosystem, we can let the auctioneer directly compute the XOR gate over the ciphertexts without knowing the private key and the input bits. Then the question is if we can also compute the AND gate privately with the Goldwasser-Micali cryptosystem. The answer is yes. Our main idea is to introduce some appropriate interactions between the auctioneer and the agent, in addition to the application of the XOR homomorphism. We now describe how to compute an AND gate privately.
\\ \indent Let $a,b\in \{0,1\}$  denote the two input bits to the AND gate, respectively. Suppose that the auctioneer holds the encrypted bits $E(a)$ and $E(b)$, and wants to compute $E(a \wedge b)$ ($\wedge$ represents AND gate) without knowing anything about the secret key and the input bits $a$ and $b$ with some interactions with the agent. The computation proceeds as follows.
\\ \indent Firstly, the auctioneer selects two random bits $r_1,r_2\in \{0,1\}$ and then calculates $E(r_1 )$ and $E(r_2)$ using the public key.
\\ \indent Then, the auctioneer can obtain the following equations according to Goldwasser-Micali homomorphic property:
\[
E(r_1)\times E(a)=E(r_1\oplus a)
\]
\[
E(r_2)\times E(b)=E(r_2\oplus b)
\]
\indent Since $a=r_1 \oplus r_1\oplus a$ and $b=r_2\oplus r_2 \oplus b$, we can decompose the bit $a$ into the XOR of two bits $a_1=r_1$ and $a_2=r_1 \oplus a$. Similarly, the bit $b$ can be computed by $b=b_1\oplus b_2$ with $b_1=r_2,b_2=r_2\oplus b$. Therefore we have
\begin{align}
  E(a \wedge b)&=E((a_1\oplus a_2)\wedge (b_1\oplus b_2)) \notag \\
  &=E(a_1\wedge b_1 \oplus a_1\wedge b_2\oplus a_2 \wedge b_1\oplus a_2\wedge b_2)  \notag \\
  &=E(a_1\wedge b_1)\times E(a_1\wedge b_2)\times E(a_2\wedge b_1)\times E(a_2\wedge b_2)
\end{align}
\indent Distinctly, due to the auctioneer knows the value of $a_1,b_1,E(a)$ and $E(b)$, so it can compute $E(a_2 ),E(b_2 ),E(a_1 \wedge b_1)$ and $E(a_1 \wedge b_2)$:
\begin{itemize}
\item[-]{$E(a_2 )=E(a_1)\times E(a)$}
\item[-]{$E(b_2 )=E(a_2)\times E(b)$}
\item[-]{When $a_1=0$, $E(a_1  \wedge b_2)=E(0)$}
\item[-]{When $a_1=1$, $E(a_1  \wedge b_2)=E(b_2 )$}
\end{itemize}\par
\indent Similarly, the value of $E(a_2 \wedge b_1)$ can easily get. Nevertheless, the auctioneer cannot calculate by itself the value of $E(a_2  \wedge b_2)$.
\\ \indent Next, the auctioneer should send the value of $E(a_2)$ and $E(b_2)$ to the agent who manages the private key. After that the agent decrypts them to obtain the value of $a_2$ and $b_2$, and then passes the calculation result of $E(a_2  \wedge b_2)$ back to the auctioneer.
\\ \indent Finally, the auctioneer can compute the value of $E(a \wedge b)$ absolutely according to the Equation (8).
\\ \indent Using the above analysis and prove, it can draw a conclusion that information leakage will be avoided as long as the auctioneer and the agent cannot collude to seek personal interests. The private AND computation is depicted in Protocol \ref{Protocol2}, where AR and AT are the abbreviation of the auctioneer and the agent
respectively.

\begin{algorithm}[H]
\floatname{algorithm}{Protocol}
  \caption{  Product of Two Bits}
  \label{Protocol2}
  \begin{algorithmic}[1]
  \Require AR holds $E(a)$ and $E(b)$
    \Ensure AR obtain $E(a\wedge b)$
    \State {\bfseries Step AR}
       \State $a_1,b_1\in \{0,1\}$; //Select randomly
      \State $E(a_2)\leftarrow E(a_1)\times E(a)$;
       \State $E(b_2)\leftarrow E(b_1)\times E(b)$;
       \If{$a_1=0$}
       \State $E(a_1\wedge b_2)\leftarrow E(0)$
       \Else
       \State $E(a_1\wedge b_2)\leftarrow E(b_2)$
       \EndIf
       \If{$b_1=0$}
       \State $E(a_2\wedge b_1)\leftarrow E(0)$
       \Else
       \State $E(a_2\wedge b_1)\leftarrow E(a_2)$
       \EndIf
       \State Sends $E(a_2)$ and $E(b_2)$ to AT;
     \State {\bfseries Step AT}
     \State $a_2\leftarrow D(E(a_2)),b_2\leftarrow D(E(b_2))$;
    \State Sends $E(a_2\wedge b_2)$ to AR;
    \State {\bfseries Step AR}
    \State $E(a \wedge b) \leftarrow E(a_1\wedge b_1)\times E(a_1\wedge b_2 )\times E(a_2\wedge b_1)\times E(a_2 \wedge b_2 )$
  \end{algorithmic}
\end{algorithm}

\subsection{Private Integer Comparison}
\begin{figure}[!htb]
\centering
\begin{minipage}[t]{0.48\textwidth}
\centering
\includegraphics[width=6cm]{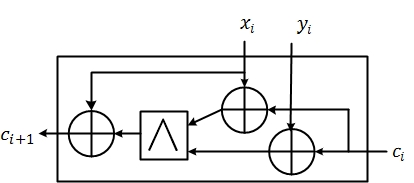}
\caption{1-bit comparator (\citealp{Kolesnikov2009Improved})}
\label{fig2}
\end{minipage}
\begin{minipage}[t]{0.48\textwidth}
\centering
\includegraphics[width=6cm]{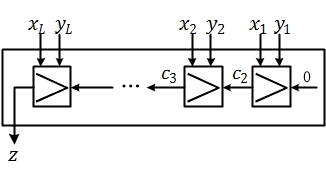}
\caption{Comparison Circuit (CMP) (\citealp{Kolesnikov2009Improved})}
\label{fig3}
\end{minipage}
\end{figure}

\indent As we can see clearly from Protocol \ref{Protocol1}, one of the most important steps is private integer comparison. Based on the private gate computation described in Section 3.2, in order to design a private protocol for integer comparison, we need only to design a Boolean circuit comprised of only XOR and AND gates, then for each gate apply either XOR or AND private gate computation step, and finally all these steps are composed as a whole protocol.
Specifically, we could take cognizance of the difference between the XOR and AND private gate computations. The former only requires computation using Equation (2) without any communication, which needs a small price; however, the cost of the latter is high by using Protocol \ref{Protocol2} in comparison with the former due to both computation and communication overheads. Just for this reason, we demand to minimize the number of AND gates as possible as we can.
\\ \indent Concretely, the procedures to design a comparison circuit and the pseudo-codes are explained as follows.
\\ \indent We denote two $L$-bits integers by $x=(x_L,x_{L-1},\cdots,x_1)$ and $y=(y_L,y_{L-1},\cdots,y_1)$. One 2-input AND gate and three XOR gates can achieve a 1-bit comparator shown in Fig. \ref{fig2}, and then $L$ sequential 1-bit comparator can compose comparison circuit (CMP as illustrated in Fig. \ref{fig3}). Specifically,
\begin{itemize}
\item[-]{For $c_1=0$, the comparison result $c_{i+1}= [x > y]$;}
\item[-]{For $c_1=1, c_{i+1}= [x \geq y]$.}
\end{itemize}\par
\indent Through comparing $x_i$ with $y_i$ from the low to the high, and then passing the result as an input to the next comparator, we achieve a private comparison of two $L$-bits positive integers $x,y$ and obtain the result $z$.
$$z=[x>y]=
\begin{cases}
1,& \text{$x > y$}\\
0,& \text{$x\leq y$}
\end{cases}$$
\\ \indent After detailed investigation and research, our design determines to apply the circuit proposed in \citet{Kolesnikov2009Improved} for integer comparison, which is optimized by ``free-XOR'' gates and exactly matches our design goals.

\begin{algorithm}[H]
\floatname{algorithm}{Algorithm}
  \caption{ Integer Comparison}
  \label{Algorithm3}
  \begin{algorithmic}[1]
  \Require $x,y$
    \Ensure Comparison result $z$
    \Function {CMP}{$x,y$}
      \State $c_1 \gets 0$
      \For{$i=1\cdots L$}
      \State $ c_{i+1} \gets x_i\oplus((x_i\oplus c_i)\wedge (y_i\oplus c_i))$
      \EndFor
      \State \Return{$z=c_{L+1}$}
    \EndFunction
  \end{algorithmic}
\end{algorithm}

\subsection{Private Sorting}
\begin{figure}
  \centering
  \subfigure[Compare-and-swap]{
    \includegraphics[width=2.0in]{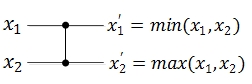}}
  \hspace{1in}
  \subfigure[Sorting Network]{
    \includegraphics[width=1.5in]{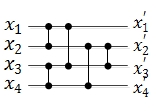}}
  \caption{Sorting Network: n = 4}
  \label{fig4} %% label for entire figure
\end{figure}

\indent As stated previously, in our main protocol we need to sort the encrypted buyers' and sellers' bid values in a certain way. Similarly to the design of private integer comparison, we can just design a sorting Boolean circuit composed of only XOR and AND gates. Based on the integer comparison circuit presented in Section 3.3, we can more easily describe the sorting circuits used in our work.
\\ \indent Due to our private mechanism design from the circuit level, our main protocol can adopt various sorting circuits. We compare the influence of different sorting circuits on our privacy-preserving auction mechanism. Specifically, we have employed the selection sort (SeSort), the bitonic sorting network (BiSort) and the odd-even merge sorting network (OESort) (\citealp{chen2016On}; \citealp{Wang2010Bureaucratic}; \citealp{Ye2010High}). The comparison result displays the immense advantage of the odd-even sorting network which is shown in Section 4 in some detail. The sorting algorithms are presented as follows.
\begin{itemize}
\item {\textbf {SeSort}. The selection sorting algorithm is noted for its simplicity which has $O(n^2)$, and it has performance advantages over more complicated algorithms in certain situations, particularly where auxiliary memory is limited.}
\item {\textbf {Sorting Network}. A circuit that sorts an input sequence $(x_1,x_2,\ldots,x_n)$ into a monotonically increasing sequence $(x'_1,x'_2,\ldots,x'_n)$. The main component of sorting network is compare-and-swap circuits, a binary operator taking as input a pair $(x_1,x_2)$, and returning the sorted pair $(x'_1,x'_2)$, with $x'_1$=$\min(x_1,x_2)$ and $x'_2$=$\max(x_1,x_2)$. Fig. $\ref{fig4}$ gives an illustration of a compare-and-swap circuit and a sorting network for $n = 4$.}
\begin{enumerate}[\indent (1)]
\item BiSort. Bitonic sort is one of the fastest sorting networks by taking the divide-and-conquer strategy. The key building blocks of a bitonic sorting network are the bitonic merge networks which rearrange bitonic sequences to be ordered. A bitonic sequence is a sequence with $x_1\leq \cdots \leq x_k\geq \cdots \geq x_n$ for some $k(1\leq k\leq n)$, or a circular shift of such a sequence.
\item OESort. The odd-even merge sorting algorithm is based on a merge algorithm that merges two sorted halves of a sequence to a completely sorted sequence. As desirable properties, the odd-even merge sorting network is data-independent and achieves a computation complexity of $O(n\log ^2 n)$. In practice, it outperforms most widely used sorting network algorithms (\citealp{Emmadi2015Updates}).
\end{enumerate}
\end{itemize}

\begin{algorithm}[H]
\floatname{algorithm}{Algorithm}
  \caption{Select Sorting}
  \label{Algorithm4}
  \begin{algorithmic}[1]
  \Require sequence $a=\{ a_0, \cdots, a_{n-1}\}$
    \Ensure the sorted sequence(non-decreasing)
    \Function {SeSort}{$a$}
    \For{$i=0\ldots n-1$}
    \State $k\leftarrow i$
    \For{$j=i+1\ldots n-1$}
      \If{$CMP(a_j,a_k)$}
      \State $k\leftarrow j$
       \EndIf
    \EndFor
    \If{$i\neq k$}
      \State $temp\leftarrow a_i$
      \State $a_i\leftarrow a_k$
      \State $a_k\leftarrow temp$
       \EndIf
    \EndFor
    \EndFunction
  \end{algorithmic}
\end{algorithm}

\begin{algorithm}[H]
 \floatname{algorithm}{Algorithm}
  \caption{Bitonic Sorting network}
  \label{Algorithm5}
  \begin{algorithmic}[1]
  \Require sequence $a=\{ a_0, \cdots, a_{n-1}\}$
    \Ensure the sorted sequence(non-decreasing)
    \Function {BiSort}{$up,a$}
    \If{$n\leq 1$}
      \State return $a$
      \Else
      \State $m \leftarrow n/2$
      \State $first\leftarrow BiSort(True,a[0\ldots m-1])$
      \State $second\leftarrow BiSort(False,a[m\ldots n-1])$
      \State return $BiMerge(up,first+second)$
      \EndIf
    \EndFunction
    \Function {BiMerge}{$up,a$}
    \If{$n\leq 1$}
      \State return $a$
      \Else
      \State $BiCMP(up,a)$
      \State $m \leftarrow n/2$
      \State $first\leftarrow BiMerge(up,a[0\ldots m-1])$
      \State $second\leftarrow BiMerge(up,a[m\ldots n-1])$
      \State return $first+second$
      \EndIf
    \EndFunction
     \Function {BiCMP}{$up,a$}
    \State $m\leftarrow n/2$
    \For{$i=0\ldots m-1$}
    \If {$CMP(a_i,a_{i+m})=true$}
    \State $a_i\Longleftrightarrow a_{i+m}$
    \EndIf
    \EndFor
    \EndFunction
  \end{algorithmic}
\end{algorithm}

\begin{algorithm}[H]
 \floatname{algorithm}{Algorithm}
  \caption{Odd-even Merge Sorting}
  \label{Algorithm6}
  \begin{algorithmic}[1]
  \Require sequence $a=\{ a_0, \cdots, a_{n-1}\}$($n$ a power of 2)
    \Ensure the sorted sequence(non-decreasing)
    \Function {OESort}{$a$}
      \If{$n>1$}
      \State apply OESort($n$/2) recursively to the two halves $a_0, ..., a_{n/2-1}$ and $a_{n/2}, ..., a_{n-1}$ of the sequence
       \State OEMerge($n$)
       \EndIf
    \EndFunction
  \end{algorithmic}
\end{algorithm}

\begin{algorithm}[H]
 \floatname{algorithm}{Algorithm}
  \caption{Odd-even Merge}
  \label{Algorithm7}
  \begin{algorithmic}[1]
  \Require sequence $a=\{ a_0, \cdots, a_{n-1}\}$ of length $n>1$ whose two halves $a_0, ..., a_{n/2-1}$ and $a_{n/2}, ..., a_{n-1}$ are sorted ($n$ a power of 2)
    \Ensure the sorted sequence
    \Function {OEMerge}{$n$}
      \If{$n>2$}
      \State apply OEMerge($n$/2) recursively to the even subsequence $a_0,a_2,..., a_{n-2}$ and $a_1,a_3,..., a_{n-1}$ of the sequence
      \For{$i=1$;$i<n$;$i\leftarrow i+2$}
        \If {$CMP(a_i,a_{i+1})=true$}
        \State $a_i\Longleftrightarrow a_{i+1}$
        \EndIf
      \EndFor
      \Else
      \If {$CMP(a_0,a_1)=true$}
        \State $a_0\Longleftrightarrow a_1$
      \EndIf
       \EndIf
    \EndFunction
  \end{algorithmic}
\end{algorithm}

\subsection{Example}
\indent Finally, we use a toy example to demonstrate our auction protocol.
\\ \indent \textbf {Step1: Setup}
\\ \indent Let us assume $M=N=5$, the bid-ID pairs of sellers $(V_i^s,s_i)$ are (200,1), (500,2), (100,3), (450,4), (150,5), and likewise the ones of buyers $(V_i^b,b_i)$ are (220,1), (180,2), (400,3), (300,4), (550,5), respectively. The agent generates its secret and public keys $(sk,pk)$.
\\ \indent \textbf {Step2: Bidding}
\\ \indent Each buyer or seller encrypts its bid-ID pair in the EBV form using the public key of the agent, and submits $E_{pk}(V_i^b,b_i)=\{ E_{pk}^L(V_i^b ),E^{L'}_{pk}(b_i)\}$ or $E_{pk}(V_i^s,s_i)=\{ E_{pk}^L(V_i^s ),E^{L'}_{pk}(s_i) \}$ to the auctioneer. For example, the buyer $b_1$ submits $E_{pk}(200,1)$ and the seller $s_1$ submits $E_{pk}(220,1)$ to the auctioneer.
\\ \indent \textbf {Step3: Private sorting}
\\ \indent The auctioneer inputs the sellers' EBV bid-ID pairs and the buyers' EBV bid-ID pairs into a private sorting protocol, and gains a non-descending order of the former and a non-ascending order of the latter:
\\ \indent The sellers' EBV bid-ID pairs:
\[
E_{pk}(100,3)\leq E_{pk}(150,5)\leq E_{pk}(200,1)\leq E_{pk}(450,4)\leq E_{pk}(500,2)
\]
\indent The buyers' EBV bid-ID pairs:
\[
E_{pk}(550,5)\geq E_{pk}(400,3)\geq E_{pk}(300,4)\geq E_{pk}(220,1)\geq E_{pk}(180,2)
\]
\indent \textbf {Step4: Private winner determination}
\\ \indent The auctioneer with the assistance of the agent compares the above encrypted sellers' bids $E_{pk}^L (V_i^s)$ with the corresponding encrypted buyers' bids $E_{pk}^L(V_i^b)$ in proper sequence until the index of the last profitable transaction $k$ is found. In light of Equation (1), we can find $k=3$, that is tantamount to mean, the first two sellers and the first two buyers are the winners.
\[
k=argmax\{ E_{pk}^L(V_k^s)\leq E_{pk}^L(V_k^b)\}
\]
\indent \textbf {Step5: Outcome release}
\\ \indent Before sending the auction outcome in the EBV form for decryption, the auctioneer needs to randomly permute the EBV IDs whose index is before 3, and then submits the resulted arrays to the agent.
\\ \indent After decryption, the agent is aware of that the group of sellers $\{s_3,s_5\}$ and the group of buyers $\{b_5,b_3\}$ are the winners. And then he needs to sort the two groups in ascending order $(\{s_3,s_5\}, \{b_3,b_5\})$ and sends all decrypted data back to the auctioneer. Finally, the auctioneer charges the price of 300 from each winning buyer and pays the price of 200 for each winning seller.

\subsection{Security Analysis}
\indent Now we prove theoretically that our mechanism (Protocol \ref{Protocol1}) preserves privacy in the sense of Definition 1. Note that since all sellers and buyers take part in the auction by providing their encrypted bids and IDs, and receiving the plain auction outcome, they do not actually participate in the essential computations. We therefore need to mainly prove that the auction computations between the auctioneer and the agent preserve privacy. To prove this, we take two steps. First, we show that our gate computations (like Protocol \ref{Protocol2}) are private. Second, we show that Protocol \ref{Protocol1} is private based on the privacy of the gate computations.

\newtheorem{thm}{Theorem}
\begin{thm}Given two encrypted bits, we can privately compute the encrypted result of a XOR or AND gate in the presence of semi-honest adversaries.
\label{theorem1}
\end{thm}
\newproof{pot}{Proof}
\begin{pot}
First, it is obvious that an XOR gate can be computed privately. Let the auctioneer holds the two encrypted bits, then it can compute the encrypted result by immediately multiplying the two input ciphers. The process is private since neither the auctioneer nor the agent knows anything about the two input bits, and both their views are some ciphertexts generated with a semantically secure cryptosystem or a random key pair, which can be easily simulated.
\\ \indent Next, we show that Protocol \ref{Protocol2} computes an AND gate privately. We show that both the auctioneer's and the agent's views can be simulated. In Protocol \ref{Protocol2}, the view of the auctioneer is
\[
\{ E(a), E(b), E(a_2\wedge b_2 ), a_1, b_1\}
\]
\indent Where the first three ciphers can be simulated with arbitrary three ciphers encrypted with the same public key due to the semantic security of the homomorphic encryption, and the latter two bits are random bits and can be simulated with two random bits drawn from the same distribution.
\\ \indent The view of the agent is $\{a_2, b_2\}$, which are also two random bits and can be simulated similarly.
\\ \indent We now can conclude that Protocol \ref{Protocol2} computes an AND gate privately. Therefore, Theorem \ref{theorem1} is proved. $\Box$
\end{pot}

\begin{thm}Protocol \ref{Protocol1} is private in the presence of semi-honest adversaries.
\label{theorem2}
\end{thm}
\begin{pot}
To prove the theorem, we need only to show that we can compute every step privately, and then sequentially compose these steps getting a whole private protocol. Now we show the privacy of each step based on Theorem \ref{theorem1}.
\\ \indent Step 1 is naturally private due to the normal cryptosystem's setup.
\\ \indent Step 2 is private due to the semantic security of the used cryptosystem.
\\ \indent Step 3 can be implemented using a sorting circuit comprised of only XOR and AND gates, and the circuit can be privately computed gate by gate. Therefore, based on Theorem \ref{theorem1} and the sequential composition theorem (\citealp{Hazay2010Efficient}), this step is also private.
\\ \indent Step 4 is private similarly to Step 3.
\\ \indent In Step 5, we carefully design the outcome release, such that only the auction outcome including the IDs of winners and the clearing prices are revealed (Note that we use a random permutation for the auctioneer and the agent, respectively, so that the bid ranking information of winners is leaked to none of them). Therefore, this step is private, too.
\\ \indent With the sequential composition theorem (\citealp{Hazay2010Efficient}), we can conclude that Protocol \ref{Protocol1} is private. $\Box$
\end{pot}

\section{PERFORMANCE EVALUATION}
\begin{figure}
\centering
\subfigure{
\begin{minipage}{0.5\textwidth}
\includegraphics[width=1\textwidth]{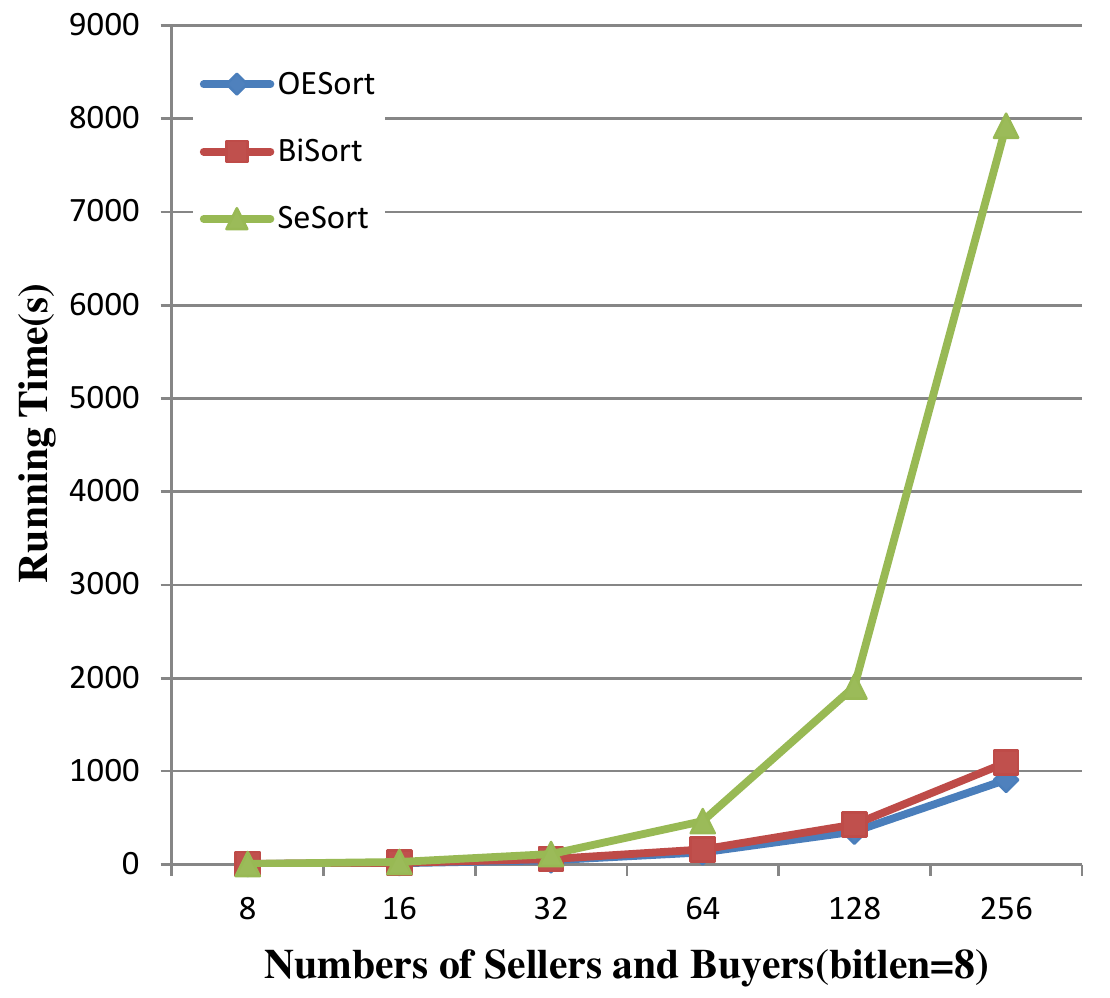}
\end{minipage}
}
\subfigure{
\begin{minipage}{0.5\textwidth}
\includegraphics[width=1\textwidth]{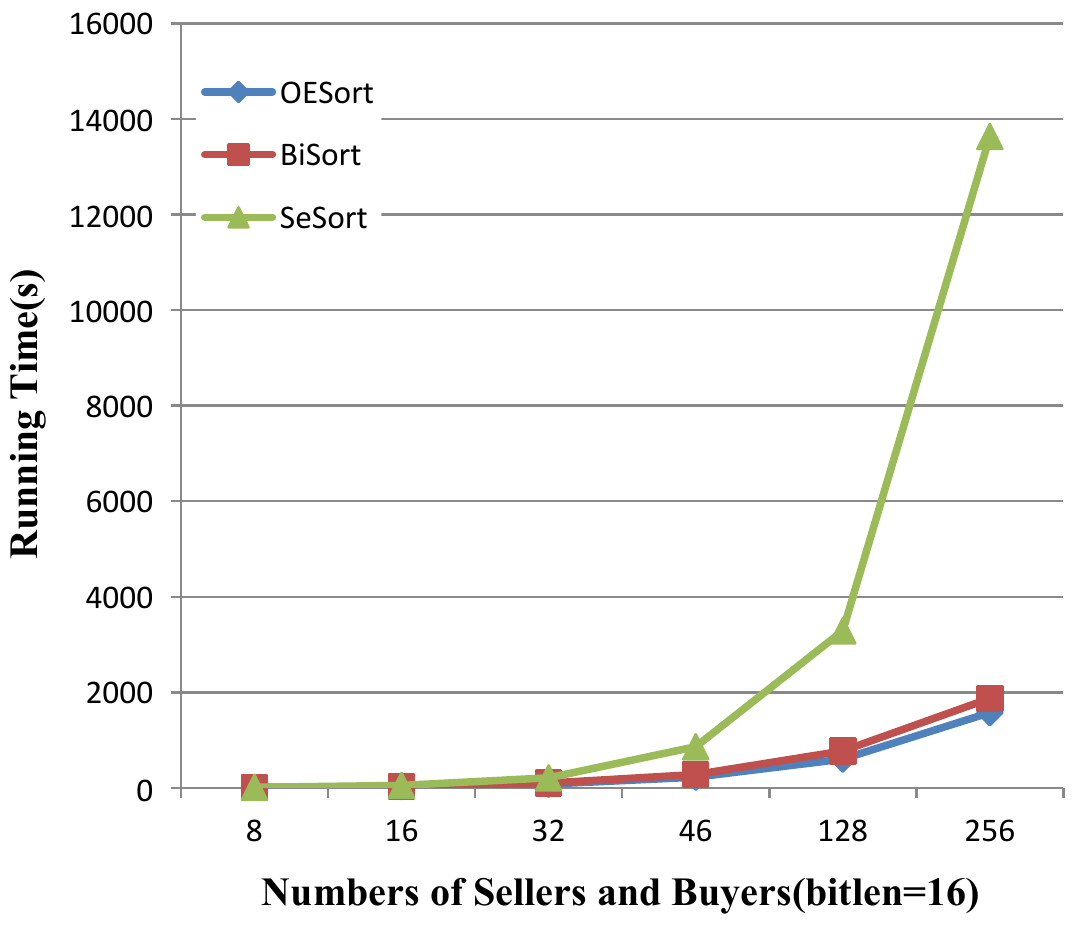}
\end{minipage}
}
 \caption{Computation comparisons for proposed protocol}
 \label{fig5}
\end{figure}

\indent In this section, we implement our proposed mechanism for double auctions using Java in Windows 10 with Intel's Core 4 Duo CPU 2.20 GHz. Furthermore, we analyze the influence of proposed protocol performance using different sorting algorithms such as the selection sort (SeSoet), the bitonic sorting network (BiSort) and the odd-even merge sorting network (OESort). We simulate the auctioneer and the agent with the two processes on a computer.
\\ \indent The experimental setting is as follows: the bit length of homomorphic encryption scheme $bitlen$ (i.e. $L$ in Definition 2) is 8 or 16 bits; the numbers $(M,N)$(assume $M=N)$ of sellers and buyers are ranging from 8 to 256 which must be a power of 2. All experimental results are averaged on 10 random repetitions. In our evaluation, we concentrate on the performance metric of computation complexity which can be reflected in total CPU time spent by the auctioneer and the agent.
\\ \indent Figure \ref{fig5} shows the curves of running times and message volumes of our proposed protocol as the numbers $M$(or $N$) vary from 8 to 256, with $bitlen$=8 and $bitlen$=16, respectively. According to the figure, we can make the following observations:
\begin{itemize}
\item{Curve Trend: For all curves, the computation complexity of our protocol is increasing as the number of sellers or buyers increases. Specially, with the growth of the number $M$(or $N$), the running time of the selection sort is rising sharply; by comparison, two other curves are rising slowly.}
\item{Performance Comparison: Given the number of sellers or buyers, the computation complexity of SeSort is much larger than that of OESort and BiSort. Obviously, the difference becomes more significant as the number of $M$(or $N$) increases. For example, when the number of sellers or buyers is 256 and $bitlen$=8, the running time of SeSort is about 132 min and that of OESort is only about 15 min.}
\item{Influence of $bitlen$: Employing the same sort algorithm, when the $bitlen$ is twice as much, the running time would increase obviously. In particular, as for the selection sort, the running time is also about double times that of the original.}
\end{itemize}\par
\indent From the analytical and experimental results above, we can see that the efficiency of the odd-even merge sorting network is significantly or slightly better than that of the selection sort or the bitonic sorting, which can achieve a computation complexity of $O(n\log ^2 n)$ and be feasible for practical applications. Take into account this reason, we are recommending the odd-even merge sorting network in the proposed protocol.

\section{RELATED WORK}
\indent In this section, we review existing work on secure and privacy-preserving auctions and mechanism designs and explore the unique features that we have proposed.
\\ \indent Double auctions have been studied extensively in recent years. For instance, Wang et al. proposed TODA (\citealp{Wang2010TODA}), a truthful online double auction for allocation in wireless networks. Samimi et al. proposed CDARA (\citealp{Samimi2016A}) to come to an optimal market-based resource allocation in cloud computing environments. Nassiri-Mofakham et al. presented a new packaging model MACBID (\citealp{Nassiri2015Bidding}) in order to find the combination (and quantities) of the items and the total price which best satisfy the bidder's need. Bing et al. investigated bidding strategies for automated traders that select marketplaces and submit offers across multiple double auction marketplaces (\citealp{Shi2016An}). However, most of them considered only single-sided auctions, and the existing double auction mechanisms do not provide too much privacy guarantee.
\\ \indent Extensive work has emphasized on privacy-preserving auction designs for conventional auctions, such as \citet{Li2015Secure}, \citet{Peng2002Robust}, \citet{Suzuki2003Secure}, and \citet{Yokoo2004Secure}. However, most of these studies (\citealp{Peng2002Robust}; \citealp{Suzuki2003Secure}; \citealp{Yokoo2004Secure}) had the drawback of confidential information disclosure in the sense of cryptography. The research \citet{Li2015Secure} did preserve privacy in the sense of cryptography like our work, but it paid more attention to double spectrum auctions making use of garbled circuits. Although utilizing garbled circuits have some advantages in terms of computation overhead, our main design targets are to achieve an alternative privacy-preserving double auction mechanism based on Goldwasser-Micali homomorphic encryption, and to analyze the influence of the proposed protocol performance with regard to different sorting algorithms. Furthermore, the protocols and algorithms proposed in our paper can also be used to combine with the garbled circuits to obtain a better performance.

\section{CONCLUSION}
\indent In this paper, we have designed an alternative privacy-preserving mechanism for double auctions. As far as we know, the proposed mechanism is the first scheme for privacy-preserving double auctions based on Goldwasser-Micali homomorphic encryption. Moreover, we have theoretically proved that the proposed mechanism preserves privacy in the presence of semi-honest adversaries. Finally, we have investigated three sorting circuits for our mechanism, and experimentally shown that different sorting algorithms may have great influence on the performance of the mechanism. We believe that our practical, easily implemented mechanism can be applied to other real-world applications for privacy protection, such as trading markets, electronic transactions and public verification of private data.



%% References

%\bibliographystyle{elsarticle-num}
\bibliographystyle{elsarticle-harv}
\bibliography{mybibfile}






% \end{thebibliography}

\end{document}